\documentclass[a4paper,aps,twocolumn,nofootinbib]{revtex4}
\RequirePackage[colorlinks,hyperindex]{hyperref}
\RequirePackage[english]{babel}
\RequirePackage[latin1]{inputenc}
\RequirePackage[T1]{fontenc}
\RequirePackage{mathrsfs}
\RequirePackage{amsmath}
\RequirePackage{amssymb}
\RequirePackage{amsbsy}
\RequirePackage{color}
\RequirePackage{bm}
\hypersetup{colorlinks=true,breaklinks=true,urlcolor=blue,linkcolor=red}
\begin{document}
%%%%%%%%%%%%%%%%%%%%%%%%%%%%%%%%%%%%%%%%%%%%%%%%%%%%%%%%%%%%%%%%%%%%%%%%%%%%%%%%%%%%%%%%%%%%%%%%%%%
\title{\bf{Symmetry Breaking of Universal type and Particular types}}
\author{Luca Fabbri\footnote{fabbri@dime.unige.it}}
\affiliation{DIME, Sez. Metodi e Modelli Matematici, Universit\`{a} di Genova,
Via all'Opera Pia 15, 16145 Genova, ITALY}
\date{\today}
%%%%%%%%%%%%%%%%%%%%%%%%%%%%%%%%%%%%%%%%%%%%%%%%%%%%%%%%%%%%%%%%%%%%%%%%%%%%%%%%%%%%%%%%%%%%%%%%%%%
\begin{abstract}
The concepts of symmetry and its breakdown are investigated in two different terms according to whether the resulting asymmetry is universal or only obtained for a special configuration: we shall illustrate this by considering in the first case an example from the standard model of particles with some consequences for cosmological scenarios related to inflation and the problem of the cosmological constant, and in the second case we consider an example from specific solutions for particle dynamics and an example for a toy model of entangled spins.
\end{abstract}
%%%%%%%%%%%%%%%%%%%%%%%%%%%%%%%%%%%%%%%%%%%%%%%%%%%%%%%%%%%%%%%%%%%%%%%%%%%%%%%%%%%%%%%%%%%%%%%%%%%
\maketitle
%%%%%%%%%%%%%%%%%%%%%%%%%%%%%%%%%%%%%%%%%%%%%%%%%%%%%%%%%%%%%%%%%%%%%%%%%%%%%%%%%%%%%%%%%%%%%%%%%%%
\section{Introduction}
In physics, one of the most important concepts is that of symmetry:\! from general coordinate covariance,\! through local Lorentz covariance, to gauge covariance, symmetry is the basis upon which all kinematical quantities can be defined. And when such kinematic quantities are coupled together into dynamic field equations, the requirement of covariance is also capable of restricting the possible ways in which this coupling can be achieved, to the point that when re-normalizability is further assumed, the possible terms within the Lagrangian are reduced to just a few.

For this reason, symmetry principles play a fundamental role. Nevertheless, starting from a theory that is symmetric, we are eventually forced to address the fact that Nature is obviously \emph{not} symmetric at all.\! Symmetry must hence be followed by a breakdown, leading to asymmetry.

Because every symmetry is represented as some invariance under transformations, symmetry breaking must be described by fixing some parameter in those transformations.\! As parameters can be either universal or proper to specific situations, it follows that there are two types of asymmetries according to whether they are obtained for the universe as a whole or for specific sub-systems. So in the first case, symmetry breaking is of the type we have in the standard model, where the Higgs vacuum $\phi^{2}\!=\!v^{2}$ is given in terms of a universal parameter,\! and in the second case,\! symmetry breaking is of the type we have when looking for solutions of field equations, where a choice of boundary conditions is different for different solutions.

In the following, we will review some of the symmetries and their breaking both for the case of universal type and for the case of particular type, and in the last case we will deepen the discussion presenting two distinct examples.
%%%%%%%%%%%%%%%%%%%%%%%%%%%%%%%%%%%%%%%%%%%%%%%%%%%%%%%%%%%%%%%%%%%%%%%%%%%%%%%%%%%%%%%%%%%%%%%%%%%
%%%%%%%%%%%%%%%%%%%%%%%%%%%%%%%%%%%%%%%%%%%%%%%%%%%%%%%%%%%%%%%%%%%%%%%%%%%%%%%%%%%%%%%%%%%%%%%%%%%
\section{Symmetry Breaking of Universal type: the Standard Model}
We will begin the treatment by recalling the Standard Model in a manner that is slightly different from the way it is usually presented, so to highlight specific features of interest. Because the SM is the theory that is symmetric under the group $\mathrm{SU(2)\!\times\!U(1)}$ we start giving generalities about these transformations and their properties.\footnote{We have assumed the reader familiar with the Pauli matrices.}

In the most general form $\mathrm{SU(2)}$ transformations are
\begin{eqnarray}
\boldsymbol{U}\!=\!e^{-\frac{i}{2}\vec{\boldsymbol{\sigma}}\cdot\vec{\theta}}
\end{eqnarray}
so that defining
\begin{eqnarray}
y^{2}\!=\!\vec{\theta}/2\!\cdot\!\vec{\theta}/2
\end{eqnarray}
and hence
\begin{eqnarray}
X\!=\!\cos{y}\\
\vec{Z}\!=\!\frac{1}{2}\frac{\sin{y}}{y}\vec{\theta}
\end{eqnarray}
we can write them as
\begin{eqnarray}
\boldsymbol{U}\!=\!X\mathbb{I}\!-\!i\vec{Z}\!\cdot\!\vec{\boldsymbol{\sigma}}
\end{eqnarray}
in explicit form. The most general form of $\mathrm{SU(2)\!\times\!U(1)}$ transformations is then given by the product of the above times a unitary phase, and since they commute we have that they can be written according to
\begin{eqnarray}
\boldsymbol{S}\!=\!\boldsymbol{U}e^{\frac{i}{2}\alpha}
\!=\!(X\mathbb{I}\!-\!i\vec{Z}\!\cdot\!\vec{\boldsymbol{\sigma}})e^{\frac{i}{2}\alpha}
\label{su(2)xu(1)}
\end{eqnarray}
as it is obvious. The doublet of complex scalar fields that transforms according to the transformation
\begin{eqnarray}
&\Phi\!\rightarrow\!\boldsymbol{S}\Phi
\end{eqnarray}
is what is known as Higgs field. From it we can build
\begin{eqnarray}
&\Phi^{\dagger}\vec{\boldsymbol{\sigma}}\Phi\!=\!\vec{S}\\
&\Phi^{\dagger}\Phi\!=\!P
\end{eqnarray}
which are both real quantities and such that
\begin{eqnarray}
&\Phi\Phi^{\dagger}\!=\!\frac{1}{2}P\mathbb{I}\!+\!\frac{1}{2}\vec{S}\!\cdot\!\vec{\boldsymbol{\sigma}}
\end{eqnarray}
as well as
\begin{eqnarray}
&\vec{S}\!\cdot\!\vec{S}\!=\!P^{2}
\end{eqnarray}
valid as general geometric identities.\footnote{These are just the Fierz identities.}

Because the Higgs field transforms in this way, we can prove that one can always find a gauge in which
\begin{eqnarray}
\Phi\!=\!\phi\boldsymbol{R}^{-1}\left(\begin{tabular}{c}
$0$\\ 
$1$
\end{tabular}\right)
\end{eqnarray}
for some $\boldsymbol{R}$ and in terms of $\phi$ being a generic real scalar field,\! and the only degree of freedom,\! called module. Then
\begin{eqnarray}
&\vec{S}\!=\!\phi^{2}\vec{s}
\end{eqnarray}
and
\begin{eqnarray}
&P\!=\!\phi^{2}
\end{eqnarray}
such that
\begin{eqnarray}
&\Phi\Phi^{\dagger}\!=\!\frac{1}{2}\phi^{2}\left(\mathbb{I}\!+\!\vec{s}\!\cdot\!\vec{\boldsymbol{\sigma}}\right)
\end{eqnarray}
as well as
\begin{eqnarray}
&\vec{s}\!\cdot\!\vec{s}\!=\!1
\end{eqnarray}
is the normalized vector of isospin. In its polar form, the Higgs $4$ real components are re-arranged into the special configuration for which the real scalar degree of freedom is isolated from the $3$ real components that are passed on into the gauge through the $3$ parameters of the $\boldsymbol{R}$ matrix that are known with the name of Goldstone bosons \cite{Goldstone:1961eq,Goldstone:1962es}.

It can be seen from expression (\ref{su(2)xu(1)}) that by introducing
\begin{eqnarray}
(\partial_{\mu}X\vec{Z}-X\partial_{\mu}\vec{Z})\!+\!\vec{Z}\!\times\!\partial_{\mu}\vec{Z}
\!=\!-\frac{1}{2}\partial_{\mu}\vec{\zeta}
\end{eqnarray}
we can always write 
\begin{eqnarray}
\boldsymbol{S}^{-1}\partial_{\mu}\boldsymbol{S}\!=\!-\frac{i}{2}\partial_{\mu}\vec{\zeta}
\!\cdot\!\vec{\boldsymbol{\sigma}}\!+\!\frac{i}{2}\partial_{\mu}\alpha\mathbb{I}\label{auxS}
\end{eqnarray}
as a general identity. Upon the introduction of two gauge fields $\vec{A}_{\mu}$ and $B_{\mu}$ as what transforms according to
\begin{eqnarray}
&g\vec{A}_{\mu}\!\cdot\!\vec{\boldsymbol{\sigma}}\!\rightarrow\!\boldsymbol{U}
\left[\left(g\vec{A}_{\mu}\!-\!\partial_{\mu}\vec{\zeta}\right)
\!\cdot\!\vec{\boldsymbol{\sigma}}\right]\boldsymbol{U}^{-1}\label{A}\\ 
&g'B_{\mu}\!\rightarrow\!g'B_{\mu}\!-\!\partial_{\mu}\alpha\label{B}
\end{eqnarray}
we can see that
\begin{eqnarray}
&D_{\mu}\Phi\!=\!\nabla_{\mu}\Phi\!-\!\frac{i}{2}\left(g\vec{A}_{\mu}
\!\cdot\!\vec{\boldsymbol{\sigma}}\!-\!g'B_{\mu}\mathbb{I}\right)\Phi
\end{eqnarray}
is the gauge covariant derivative of the Higgs field \cite{Weinberg:1967tq}.

Since from the polar form of the Higgs we can see that
\begin{eqnarray}
\boldsymbol{R}^{-1}\partial_{\mu}\boldsymbol{R}
\!=\!-\frac{i}{2}\partial_{\mu}\vec{\xi}\!\cdot\!\vec{\boldsymbol{\sigma}}
\!+\!\frac{i}{2}\partial_{\mu}\xi\mathbb{I}\label{auxR}
\end{eqnarray}
where $\xi$ and $\vec{\xi}$ are the Goldstone modes, we can define
\begin{eqnarray}
&g\vec{M}_{\mu}\!=\!g\vec{A}_{\mu}\!-\!\partial_{\mu}\vec{\xi}\label{M}\\ 
&g'N_{\mu}\!=\!g'B_{\mu}\!-\!\partial_{\mu}\xi\label{N}
\end{eqnarray}
which are proven to be true vector fields, from which
\begin{eqnarray}
&D_{\mu}\Phi\!=\!\left[\nabla_{\mu}\ln{\phi}
\!-\!\frac{i}{2}\left(g\vec{M}_{\mu}\!\cdot\!\vec{\boldsymbol{\sigma}}
\!-\!g'N_{\mu}\mathbb{I}\right)\right]\Phi
\end{eqnarray}
is the gauge covariant derivative of the Higgs field and
\begin{eqnarray}
&\nabla_{\mu}\vec{s}\!=\!g\vec{M}_{\mu}\!\times\!\vec{s}
\end{eqnarray}
are general identities. The Goldstone states are absorbed by the gauge fields as their longitudinal components \cite{Fabbri:2021mfc}.

As for the dynamics, we will consider the Lagrangian
\begin{eqnarray} 
&\mathscr{L}\!=\!D_{\mu}\Phi^{\dagger}D^{\mu}\Phi
\!-\!\frac{1}{2}\lambda^{2}\!\left(v^{2}\!-\!\Phi^{\dagger}\Phi\right)^{2}
\end{eqnarray}
with the $v$ and $\lambda$ constants and $\mathrm{SU(2)\!\times\!U(1)}$ invariant.\footnote{This potential is slightly different from the usual one for reasons that will become apparent later on.}

Plugging the polar form of the gauge covariant derivative we obtain the polar form of the Lagrangian
\begin{eqnarray}
\nonumber
&\mathscr{L}\!=\!\nabla_{\mu}\phi\nabla^{\mu}\phi
\!+\!\frac{1}{4}\phi^{2}(g^{2}\vec{M}_{\mu}\!\cdot\!\vec{M}^{\mu}\!+\!g'^{2}N^{\mu}N_{\mu}-\\
&-2gg'N^{\mu}\vec{M}_{\mu}\!\cdot\!\vec{s})
\!-\!\frac{1}{2}\lambda^{2}\!\left(v^{2}\!-\!\phi^{2}\right)^{2}
\end{eqnarray}
where we have no information about the direction of the isospin. If it is along the third axis of the internal space
\begin{eqnarray}
&-g\vec{M}_{\mu}\!\cdot\!\vec{s}\!=\!gM_{\mu}^{3}
\end{eqnarray}
and the Lagrangian becomes
\begin{eqnarray} 
\nonumber
&\mathscr{L}\!=\!\nabla_{\mu}\phi\nabla^{\mu}\phi
\!+\!\frac{1}{4}\phi^{2}[g^{2}(M_{\mu}^{1}M^{\mu}_{1}\!+\!M_{\mu}^{2}M^{\mu}_{2})+\\
&+(gM_{\mu}^{3}\!+\!g'N_{\mu})(gM^{\mu}_{3}\!+\!g'N^{\mu})]
\!-\!\frac{1}{2}\lambda^{2}\!\left(v^{2}\!-\!\phi^{2}\right)^{2}
\end{eqnarray}
so that after diagonalizing as
\begin{eqnarray}
&\frac{1}{\sqrt{2}}\left(M^{1}_{\mu}\pm iM^{2}_{\mu}\right)=W_{\mu}^{\pm}\\
&gM^{\mu}_{3}\!+\!g'N^{\mu}=\sqrt{g^{2}\!+\!g'^{2}}Z^{\mu}\\
&-g'M^{\mu}_{3}\!+\!gN^{\mu}=\sqrt{g^{2}\!+\!g'^{2}}A^{\mu}
\end{eqnarray}
we eventually obtain
\begin{eqnarray} 
\nonumber
&\mathscr{L}\!=\!\nabla_{\mu}\phi\nabla^{\mu}\phi
\!+\!\frac{1}{4}\phi^{2}[2g^{2}W^{+}W^{-}+\\
&+(g^{2}\!+\!g'^{2})Z^{2}]\!-\!\frac{1}{2}\lambda^{2}\!\left(v^{2}\!-\!\phi^{2}\right)^{2}
\end{eqnarray}
as it is easy to see. The potential is defined so to be zero at its minimum. It has two equilibria, an unstable one at $\phi\!=\!0$ and a stable one at $\phi^{2}\!=\!v^{2}$ exactly like in the usual version of the standard model. The universe starts with an initial symmetric configuration $\phi\!=\!0$ for which 
\begin{eqnarray}
&\mathscr{L}\!=\!-\frac{1}{2}\lambda^{2}v^{4}
\end{eqnarray}
which is a positive and large cosmological constant term, with the consequence that the Einstein gravitational field equations become $8R_{\alpha\nu}\!=\!-\lambda^{2}v^{4}g_{\alpha\nu}$ in which $R_{\alpha\nu}$ is the Ricci tensor and $g_{\alpha\nu}$ is the metric tensor. As well known, when the Ricci tensor is proportional to the metric tensor with negative proportionality constant,\! the scale factor is an exponential function of the cosmological time and an inflationary epoch might take place.\! Nevertheless, such a condition is unstable, so that it will spontaneously move toward a stable state. The stable configuration $\phi^{2}\!=\!v^{2}$ is such that in it the cosmological constant is cancelled by the vacuum of the Higgs field, quenching inflation.\! But in the new vacuum, symmetry has been broken. Therefore we can re-parametrize this new vacuum according to
\begin{eqnarray}
&\phi\!=\!v\!+\!H
\end{eqnarray}
and consequently
\begin{eqnarray} 
\nonumber
&\mathscr{L}\!=\!\nabla_{\mu}H\nabla^{\mu}H\!-\!\frac{1}{4}m_{H}^{2}H^{4}/v^{2}
\!-\!m_{H}^{2}H^{3}/v\!-\!m_{H}^{2}H^{2}+\\
\nonumber
&+(H^{2}/v^{2}\!+\!2H/v)(m_{W}^{2}W^{+}W^{-}\!+\!\frac{1}{2}m_{Z}^{2}Z^{2})+\\
&+(m_{W}^{2}W^{+}W^{-}\!+\!\frac{1}{2}m_{Z}^{2}Z^{2})
\end{eqnarray}
where we have set $2\lambda^{2}v^{2}\!=\!m_{H}^{2}$ as the Higgs mass together with $g^{2}v^{2}\!=\!2m_{W}^{2}$ and $v^{2}(g^{2}\!+\!g'^{2})\!=\!2m_{Z}^{2}$ as the two vector boson masses like it is done in the usual standard model.

The known phenomenology, apart from the dynamical structure of the Lagrangian, is obtained by the conditions $s_{a}\!=\!-(0,0,1)$ and $\phi\!=\!v\!+\!H$ which together form the full symmetry breaking conditions. Both are universal in the sense that they are a choice of fields and parameters that is the same throughout the whole cosmological setting.

We will next observe that this is not always the case.
%%%%%%%%%%%%%%%%%%%%%%%%%%%%%%%%%%%%%%%%%%%%%%%%%%%%%%%%%%%%%%%%%%%%%%%%%%%%%%%%%%%%%%%%%%%%%%%%%%%
%%%%%%%%%%%%%%%%%%%%%%%%%%%%%%%%%%%%%%%%%%%%%%%%%%%%%%%%%%%%%%%%%%%%%%%%%%%%%%%%%%%%%%%%%%%%%%%%%%%
\section{Symmetry Breaking of Particular type: Spinorial Fields}
We continue the presentation introducing the concept of tetrads, or frames, in the general theory of relativity.\footnote{It is important to specify from the start that here with general relativity we do not mean the dynamical theory of Einsteinian gravity, obtained by assigning the field equations that link the space-time curvature to the energy tensor and interpreting the space-time curvature as gravitation. Here with general relativity we mean the kinematical theory that implements the principle of general covariance under curvilinear coordinate transformations by employing tensor quantities. As such we are simply meaning to retain the possibility to study fields in whatever system of reference, whether or not the space-time has a curvature.}

The initial point is that of assuming the existence of a metric tensor $g_{\mu\nu}\!=\!g_{\nu\mu}$ with inverse $g^{\mu\nu}\!=\!g^{\nu\mu}$ such that we have $g_{\mu\rho}g^{\rho\alpha}\!=\!\delta^{\alpha}_{\mu}$ where $\delta^{\alpha}_{\mu}$ is the Kronecker delta. We specify that this metric need not be in Minkowskian form because we retain the right to employ whatever system of coordinates, even in flat space-times. Nonetheless, we can always introduce bases of vectors $e^{\mu}_{a}$ and $e_{\mu}^{a}$ such that
\begin{eqnarray}
e^{\mu}_{a}e_{\mu}^{b}\!=\!\delta^{b}_{a}\ \ \ \ \ \ \ \ 
e^{\mu}_{a}e_{\nu}^{a}\!=\!\delta^{\mu}_{\nu}
\end{eqnarray}
called tetrads and for which 
\begin{eqnarray}
e_{\mu}^{a}e_{\nu}^{b}g^{\mu\nu}\!=\!\eta^{ab}\ \ \ \ 
e^{\mu}_{a}e^{\nu}_{b}g_{\mu\nu}\!=\!\eta_{ab}
\end{eqnarray}
where $\eta$ is the Minkowskian matrix. That we can always choose to do this comes from the fact that we can always make the ortho-normalization procedure on the tetradic fields. The tetradic fields have two indices, the Latin one indicates what vector of the basis we choose, the Greek one denotes what component of that vector we pick. Like for the metric, the Greek index is associated to a general coordinate transformation. The Latin index is associated to a new type of transformation shuffling vectors within the basis and consequently this transformation must be a Lorentz transformation since we wish the Minkowskian matrix to be preserved. By indicating such a real Lorentz transformation as $\Lambda$ we have that it acts according to
\begin{eqnarray}
e_{\nu}^{a}\!\rightarrow\!(\Lambda)^{a}_{b}e_{\nu}^{b}\ \ \ \ \ \ \ \ \ \ \ \ \ \ \ \ 
e^{\nu}_{a}\!\rightarrow\!(\Lambda^{-1})^{b}_{a}e^{\nu}_{b}
\end{eqnarray}
as the transformation on tetrads.\! We also introduce a set of Clifford matrices $\boldsymbol{\gamma}^{a}$ verifying the relations
\begin{eqnarray}
&\{\boldsymbol{\gamma}^{a},\boldsymbol{\gamma}^{b}\}\!=\!2\mathbb{I}\eta^{ab}
\end{eqnarray}
where $\mathbb{I}$ is the identity matrix. It is then possible to define 
\begin{eqnarray}
&\frac{1}{4}\left[\boldsymbol{\gamma}^{a},\boldsymbol{\gamma}^{b}\right]
\!=\!\boldsymbol{\sigma}^{ab}
\end{eqnarray}
and one can easily verify that the $\boldsymbol{\sigma}^{ab}$ thus defined satisfy the commutation relations defining the complex Lorentz algebra and therefore they are the generators of the complex Lorentz group. We also have that
\begin{eqnarray}
&2i\boldsymbol{\sigma}_{ab}\!=\!\varepsilon_{abcd}\boldsymbol{\pi}\boldsymbol{\sigma}^{cd}
\end{eqnarray}
implicitly defining the $\boldsymbol{\pi}$ matrix, whose existence proves that the complex Lorentz group is reducible. From them
\begin{eqnarray}
&\boldsymbol{\gamma}_{i}\boldsymbol{\gamma}_{j}\boldsymbol{\gamma}_{k}
\!=\!\boldsymbol{\gamma}_{i}\eta_{jk}-\boldsymbol{\gamma}_{j}\eta_{ik}
\!+\!\boldsymbol{\gamma}_{k}\eta_{ij}
\!+\!i\varepsilon_{ijkq}\boldsymbol{\pi}\boldsymbol{\gamma}^{q}
\end{eqnarray}
which are valid as geometric identities.\footnote{This matrix is usually denoted as a gamma matrix with an index five. As this index has no sense in four-dimensional space-times we will adopt a notation without any index at all.}

In the most general form the complex Lorentz transformation is written according to
\begin{eqnarray}
&\boldsymbol{\Lambda}\!=\!e^{\frac{1}{2}\theta_{ab}\boldsymbol{\sigma}^{ab}}
\end{eqnarray}
so that defining
\begin{eqnarray}
a\!=\!-\frac{1}{8}\theta_{ij}\theta^{ij}\\
b\!=\!\frac{1}{16}\theta_{ij}\theta_{ab}\varepsilon^{ijab}
\end{eqnarray}
and then
\begin{eqnarray}
2x^{2}\!=\!a\!+\!\sqrt{a^{2}\!+\!b^{2}}\\
2y^{2}\!=\!-a\!+\!\sqrt{a^{2}\!+\!b^{2}}
\end{eqnarray}
we can introduce
\begin{eqnarray}
\cos{y}\cosh{x}\!=\!X\\
\sin{y}\sinh{x}\!=\!Y\\
\nonumber
\left(\frac{x\sinh{x}\cos{y}
+y\sin{y}\cosh{x}}{x^{2}+y^{2}}\right)\theta^{ab}+\\
+\left(\frac{x\cosh{x}\sin{y}
-y\cos{y}\sinh{x}}{x^{2}+y^{2}}\right)\!\frac{1}{2}\theta_{ij}\varepsilon^{ijab}\!=\!Z^{ab}
\end{eqnarray}
so to write
\begin{eqnarray}
\boldsymbol{\Lambda}\!=\!
X\mathbb{I}\!+\!Yi\boldsymbol{\pi}+\frac{1}{2}Z^{ab}\boldsymbol{\sigma}_{ab}
\end{eqnarray}
in the most compact and explicit way.\! The most complete complex Lorentz and phase transformation is therefore
\begin{eqnarray}
&\boldsymbol{S}\!=\!\boldsymbol{\Lambda}e^{iq\alpha}\!=\!
(X\mathbb{I}\!+\!Yi\boldsymbol{\pi}\!+\!\frac{1}{2}Z^{ab}\boldsymbol{\sigma}_{ab})e^{iq\alpha}
\label{sl(2)xsl(2)}
\end{eqnarray}
called spinorial transformation. Any column of $4$ complex functions that, under general coordinate transformations, are scalars but which, for a spinorial transformation, are converted according to the law
\begin{eqnarray}
&\psi\!\rightarrow\!\boldsymbol{S}\psi
\end{eqnarray}
is called spinorial field. It is possible to prove that with the adjoint $\overline{\psi}\!=\!\psi^{\dagger}\boldsymbol{\gamma}^{0}$ we can construct the quantities
\begin{eqnarray}
&\Sigma^{ab}\!=\!2\overline{\psi}\boldsymbol{\sigma}^{ab}\boldsymbol{\pi}\psi\\
&M^{ab}\!=\!2i\overline{\psi}\boldsymbol{\sigma}^{ab}\psi\\
&S^{a}\!=\!\overline{\psi}\boldsymbol{\gamma}^{a}\boldsymbol{\pi}\psi\\
&U^{a}\!=\!\overline{\psi}\boldsymbol{\gamma}^{a}\psi\\
&\Theta\!=\!i\overline{\psi}\boldsymbol{\pi}\psi\\
&\Phi\!=\!\overline{\psi}\psi
\end{eqnarray}
which are all real tensors and such that
\begin{eqnarray}
\nonumber
&\psi\overline{\psi}\!\equiv\!\frac{1}{4}\Phi\mathbb{I}
\!+\!\frac{1}{4}U_{a}\boldsymbol{\gamma}^{a}
\!+\!\frac{i}{8}M_{ab}\boldsymbol{\sigma}^{ab}-\\
&-\frac{1}{8}\Sigma_{ab}\boldsymbol{\sigma}^{ab}\boldsymbol{\pi}
\!-\!\frac{1}{4}S_{a}\boldsymbol{\gamma}^{a}\boldsymbol{\pi}
\!-\!\frac{i}{4}\Theta \boldsymbol{\pi}\label{Fierz}
\end{eqnarray}
as well as
\begin{eqnarray}
&\Sigma^{ab}\!=\!-\frac{1}{2}\varepsilon^{abij}M_{ij}
\end{eqnarray}
with
\begin{eqnarray}
&M_{ab}\Theta\!+\!\Sigma_{ab}\Phi\!=\!U_{[a}S_{b]}
\end{eqnarray}
alongside to
\begin{eqnarray}
&U_{a}S^{a}\!=\!0\label{orthogonal1}\\
&U_{a}U^{a}\!=\!-S_{a}S^{a}\!=\!\Theta^{2}\!+\!\Phi^{2}\label{norm1}
\end{eqnarray}
as is straightforward to prove and called Fierz identities.

If we are in the situation in which $\Theta^{2}\!+\!\Phi^{2}\!\neq\!0$ then we have that it is always possible to write the spinor as
\begin{eqnarray}
&\!\psi\!=\!\phi e^{-\frac{i}{2}\beta\boldsymbol{\pi}}
\boldsymbol{L}^{-1}\left(\begin{tabular}{c}
$1$\\
$0$\\
$1$\\
$0$
\end{tabular}\right)
\label{spinorch}
\end{eqnarray}
in chiral representation, with $\boldsymbol{L}$ a Lorentz transformation and with $\phi$ and $\beta$ that are real scalar and pseudo-scalar fields,\! and the only degrees of freedom,\! called module and Yvon-Takabayashi angle. We then can compute
\begin{eqnarray}
&S^{a}\!=\!2\phi^{2}s^{a}\\
&U^{a}\!=\!2\phi^{2}u^{a}
\end{eqnarray}
as well as
\begin{eqnarray}
&\Theta\!=\!2\phi^{2}\sin{\beta}\\
&\Phi\!=\!2\phi^{2}\cos{\beta}
\end{eqnarray}
from which
\begin{eqnarray}
&\psi\overline{\psi}\!\equiv\!\frac{1}{2}\phi^{2}e^{-i\beta\boldsymbol{\pi}}
(e^{i\beta\boldsymbol{\pi}}\!+\!u_{a}\boldsymbol{\gamma}^{a})
(e^{-i\beta\boldsymbol{\pi}}\!-\!s_{a}\boldsymbol{\gamma}^{a}\boldsymbol{\pi})
\end{eqnarray}
and
\begin{eqnarray}
&u_{a}s^{a}\!=\!0\\
&u_{a}u^{a}\!=\!-s_{a}s^{a}\!=\!1
\end{eqnarray}
are the normalized velocity vector and spin axial-vector, as well known. Written in polar form, the $8$ real components of the spinor can be rearranged in such a way that the $2$ real scalar degrees of freedom are isolated from the $6$ real components that can always be transferred into the frame through the $6$ parameters of the Lorentz transformation $\boldsymbol{L}$ which can be identified as Goldstone bosons.\footnote{The previous analysis can be performed also in the case in which $\Theta\!=\!\Phi\!\equiv\!0$ although in this case we talk about a very specific type of spinors that we are not going to treat in the present work.}

As above, it can be seen from (\ref{sl(2)xsl(2)}) that defining
\begin{eqnarray}
\nonumber
&(\partial_{\mu}XZ^{ab}-X\partial_{\mu}Z^{ab})
+\frac{1}{2}(\partial_{\mu}YZ_{ij}-Y\partial_{\mu}Z_{ij})\varepsilon^{ijab}+\\
&+\partial_{\mu}Z^{ak}Z^{b}_{\phantom{b}k}\!=\!-\partial_{\mu}\zeta^{ab}
\end{eqnarray}
allows us to write
\begin{eqnarray}
\boldsymbol{S}^{-1}\partial_{\mu}\boldsymbol{S}
\!=\!\frac{1}{2}\partial_{\mu}\zeta_{ab}\boldsymbol{\sigma}^{ab}
\!+\!iq\partial_{\mu}\alpha\mathbb{I}
\end{eqnarray}
as a general identity. With the spin connection $\Omega_{ij\mu}$ and the gauge field $A_{\mu}$ given in terms of their transformations
\begin{eqnarray}
&\frac{1}{2}\Omega_{ij\mu}\boldsymbol{\sigma}^{ij}\!\rightarrow\!\boldsymbol{\Lambda}
\left[\frac{1}{2}(\Omega_{ij\mu}\!-\!\partial_{\mu}\zeta_{ij})\boldsymbol{\sigma}^{ij}\right]\boldsymbol{\Lambda}^{-1}\\
&A_{\mu}\!\rightarrow\!A_{\mu}\!-\!\partial_{\mu}\alpha
\end{eqnarray}
we have that
\begin{eqnarray}
&\boldsymbol{\nabla}_{\mu}\psi\!=\!\partial_{\mu}\psi\!+\!\frac{1}{2}\Omega_{ij\mu}\boldsymbol{\sigma}^{ij}\psi
\!+\!iqA_{\mu}\psi
\end{eqnarray}
is the spinorial covariant derivative of the spinor field \cite{G}.

From the polar form of the spinor we can also see that
\begin{eqnarray}
\boldsymbol{L}^{-1}\partial_{\mu}\boldsymbol{L}\!=\!iq\partial_{\mu}\xi\mathbb{I}
\!+\!\frac{1}{2}\partial_{\mu}\xi^{ab}\boldsymbol{\sigma}_{ab}\label{LdL}
\end{eqnarray}
where $\xi$ and $\xi^{ab}$ are the Goldstone states, so defining
\begin{eqnarray}
&q(\partial_{\mu}\xi\!-\!A_{\mu})\!\equiv\!P_{\mu}\label{P}\\
&\partial_{\mu}\xi_{ij}\!-\!\Omega_{ij\mu}\!\equiv\!R_{ij\mu}\label{R}
\end{eqnarray}
which are true tensor fields, we have that
\begin{eqnarray}
&\!\!\!\!\!\!\!\!\boldsymbol{\nabla}_{\mu}\psi\!=\!(-\frac{i}{2}\nabla_{\mu}\beta\boldsymbol{\pi}
\!+\!\nabla_{\mu}\ln{\phi}\mathbb{I}
\!-\!iP_{\mu}\mathbb{I}\!-\!\frac{1}{2}R_{ij\mu}\boldsymbol{\sigma}^{ij})\psi
\label{decspinder}
\end{eqnarray}
as spinorial covariant derivative such that 
\begin{eqnarray}
&\nabla_{\mu}s_{i}\!=\!R_{ji\mu}s^{j}\label{ds}\\
&\nabla_{\mu}u_{i}\!=\!R_{ji\mu}u^{j}\label{du}
\end{eqnarray}
as general identities. The Goldstone states are absorbed by the gauge field and spin connection thus becoming the longitudinal components of the $P_{\mu}$ and $R_{ji\mu}$ tensors \cite{Fabbri:2021mfc}.

As for the dynamics, we consider the Dirac equations
\begin{eqnarray}
&i\boldsymbol{\gamma}^{\mu}\boldsymbol{\nabla}_{\mu}\psi
\!-\!XW_{\mu}\boldsymbol{\gamma}^{\mu}\boldsymbol{\pi}\psi\!-\!m\psi\!=\!0
\label{D}
\end{eqnarray}
with $W_{\mu}$ axial-vector torsion and $X$ torsion-spin coupling constant, added to be in the most general case \cite{G}.

In polar form these equations decompose according to
\begin{eqnarray}
&\!\!\!\!B_{\mu}\!-\!2P^{\iota}u_{[\iota}s_{\mu]}\!+\!(\nabla\beta\!-\!2XW)_{\mu}
\!+\!2s_{\mu}m\cos{\beta}\!=\!0\label{dep1}\\
&\!\!\!\!R_{\mu}\!-\!2P^{\rho}u^{\nu}s^{\alpha}\varepsilon_{\mu\rho\nu\alpha}\!+\!2s_{\mu}m\sin{\beta}
\!+\!\nabla_{\mu}\ln{\phi^{2}}\!=\!0\label{dep2}
\end{eqnarray}
with $R_{\mu a}^{\phantom{\mu a}a}\!=\!R_{\mu}$ and $\frac{1}{2}\varepsilon_{\mu\alpha\nu\iota}R^{\alpha\nu\iota}\!=\!B_{\mu}$ and which can be proven equivalent to the Dirac equations.\! In fact they are two special Gordon decompositions which, in polar form, possess the same information of the Dirac equations \cite{Fabbri:2020ypd}.

We have no information about the direction of velocity and spin, although we can always boost in the rest frame and there align the spin along the third axis. In doing so we get the possibility to choose $P_{\nu}$ and $R_{ij\nu}$ in ways that might allow us to find spinorial field solutions that could be written in the radial and angular coordinates without variable separability \cite{Fabbri:2021weq,Fabbri:2021nkn}.\! As solutions\! like this depend on the elevation angle, the spinor symmetry can never be more than an axial symmetry even if the background had spherical symmetry.\! The fact that the solution of a given equation has less symmetry than that very equation tells that symmetry breaking occurred. It is of particular type as it occurs only for that specific solution. And it occurs only for that solution due to the boundary conditions.

Similarly to the previous case, we have that situations of a given symmetry allow the system to be re-configured into a form in which some degree of freedom, recognized as Goldstone states, are transferred into gauge fields, and symmetry breaking can occur. Differently from the previous case, where the symmetry breaking meant selecting a configuration of fields, now symmetry breaking means selecting a configuration of components within a field.

However, for this last case, there are two ways. In the case just described, symmetry breaking of particular type occurs because a solution is defined in terms of boundary conditions that differ for different solutions but which are the same for a single solution. But we can also have cases where properties of a given solution are set by boundary conditions that differ even for a single solution.\! Boundary conditions are fixed for $\phi$ and $\beta$ since they are determined by the field equations. But there can be no fixing $\boldsymbol{L}$ since there is no way to determine it from any field equation.

This is what we intend to do in the next section.
%%%%%%%%%%%%%%%%%%%%%%%%%%%%%%%%%%%%%%%%%%%%%%%%%%%%%%%%%%%%%%%%%%%%%%%%%%%%%%%%%%%%%%%%%%%%%%%%%%%
%%%%%%%%%%%%%%%%%%%%%%%%%%%%%%%%%%%%%%%%%%%%%%%%%%%%%%%%%%%%%%%%%%%%%%%%%%%%%%%%%%%%%%%%%%%%%%%%%%%
\section{Symmetry Breaking of Particular type: Spin Entanglement}
The polar decomposition for spinors allows us to obtain the equivalent of the Madelung decomposition in the case of relativistic situations \cite{Yvon1940}. Thus relativistic spinors can be interpreted from the hydrodynamic perspective \cite{Takabayasi1957}.

To better see this point, consider now the polar form of Dirac equations (\ref{dep1}, \ref{dep2}) written as
\begin{eqnarray}
&Y_{\mu}\!-\!P^{\iota}u_{[\iota}s_{\mu]}\!+\!ms_{\mu}\cos{\beta}\!=\!0\\
&Z_{\mu}\!+\!P^{\rho}u^{\nu}s^{\alpha}\varepsilon_{\mu\rho\nu\alpha}
\!-\!ms_{\mu}\sin{\beta}\!=\!0
\end{eqnarray}
where $(\nabla\beta\!-\!2XW\!+\!B)_{k}\!=\!2Y_{k}$ and $(\nabla\ln{\phi^{2}}\!+\!R)_{k}\!=\!-2Z_{k}$ are potentials. So we can invert the momentum \cite{Fabbri:2019tad} as
\begin{eqnarray}
&P^{\rho}\!=\!m\cos{\beta}u^{\rho}\!+\!Y_{\nu}u^{[\nu}s^{\rho]}
\!+\!Z_{\mu}s_{\alpha}u_{\nu}\varepsilon^{\mu\alpha\nu\rho}
\label{momentum}
\end{eqnarray}
after straightforward manipulation.\! This form shows that the momentum $P_{\nu}$ is not just the kinematic momentum $mu_{\nu}$ but there are a number of corrections. One is in the correction due to the\! Yvon-Takabayashi angle\! $\cos{\beta}$ which expresses the effects of internal dynamics \cite{Fabbri:2020ypd}. The others are proportional to the spin axial-vector and due to the $Y_{\nu}$ and $Z_{\mu}$ potentials. These are given by some external contributions of $W_{\alpha}$ and $R_{ij\alpha}$ plus the derivatives of the $\beta$ and $\ln{\phi^{2}}$ and as such they can be seen as the quantum potentials in relativistic version with spin. The fact that they are first-order differential is the consequence of their relativistic essence and the existence of a second quantum potential is the consequence of the internal structure that comes from the presence of spin. Both potentials are in terms containing the spin axial-vector, and consequently they disappear in the macroscopic approximation.\footnote{That the macroscopic approximation be encoded by the condition $s_{a}\!\rightarrow\!0$ is clear from the fact that if we were not to normalize $\hbar\!=\!1$ then the spin axial-vector would be multiplied by the Planck constant, and $\hbar\!\rightarrow\!0$ is the definition of the non-quantum limit.}

The above considerations also help understanding how this formalism is the best-suited formalism in which one can set the de Broglie-Bohm theory in its relativistic form with spin. To see that, let us write the expression of the energy of the spinorial field in its polar form. We have 
\begin{eqnarray}
\nonumber
&E^{\rho\sigma}\!=\!\frac{i}{4}(\overline{\psi}\boldsymbol{\gamma}^{\rho}\boldsymbol{\nabla}^{\sigma}\psi
\!-\!\boldsymbol{\nabla}^{\sigma}\overline{\psi}\boldsymbol{\gamma}^{\rho}\psi+\\
\nonumber
&+\overline{\psi}\boldsymbol{\gamma}^{\sigma}\boldsymbol{\nabla}^{\rho}\psi
\!-\!\boldsymbol{\nabla}^{\rho}\overline{\psi}\boldsymbol{\gamma}^{\sigma}\psi)-\\
&-\frac{1}{2}X(W^{\sigma}\overline{\psi}\boldsymbol{\gamma}^{\rho}\boldsymbol{\pi}\psi
\!+\!W^{\rho}\overline{\psi}\boldsymbol{\gamma}^{\sigma}\boldsymbol{\pi}\psi)
\end{eqnarray}
which in polar form becomes
\begin{eqnarray}
\nonumber
&E^{\rho\sigma}\!=\!\phi^{2}\left[(\nabla\beta/2\!-\!XW)^{\sigma}s^{\rho}
\!+\!(\nabla\beta/2\!-\!XW)^{\rho}s^{\sigma}+\right.\\
&\left.+P^{\sigma}u^{\rho}\!+\!P^{\rho}u^{\sigma}
\!-\!\frac{1}{4}(R_{ij}^{\phantom{ij}\sigma}\varepsilon^{\rho ijk}
\!+\!R_{ij}^{\phantom{ij}\rho}\varepsilon^{\sigma ijk})s_{k}\right]
\end{eqnarray}
as it is straightforward to see. Employing (\ref{momentum}) gives
\begin{eqnarray}
\nonumber
&E^{\rho\sigma}\!=\!2\phi^{2}m\cos{\beta}u^{\sigma}u^{\rho}
\!+\!\frac{1}{2}\phi^{2}\left[2Y^{\sigma}s^{\rho}\!+\!2Y^{\rho}s^{\sigma}-\right.\\
\nonumber
&\left.-2Y_{k}(s^{[k}u^{\sigma]}u^{\rho}\!+\!s^{[k}u^{\rho]}u^{\sigma})+\right.\\
\nonumber
&\left.+2Z_{k}s_{j}u_{i}(u^{\rho}\varepsilon^{kji\sigma}
\!+\!u^{\sigma}\varepsilon^{kji\rho})-\right.\\
&\left.-B^{\sigma}s^{\rho}\!-\!B^{\rho}s^{\sigma}
\!-\!\frac{1}{2}(R_{ij}^{\phantom{ij}\sigma}\varepsilon^{\rho ijk}
\!+\!R_{ij}^{\phantom{ij}\rho}\varepsilon^{\sigma ijk})s_{k}\right]
\end{eqnarray}
which is general. In macroscopic limit $s_{j}\!\rightarrow\!0$ we get
\begin{eqnarray}
&P^{\mu}\!\approx\!mu^{\mu}\cos{\beta}
\end{eqnarray}
as well as
\begin{eqnarray}
&E^{\rho\sigma}\!\approx\!2\phi^{2}m\cos{\beta}u^{\sigma}u^{\rho}
\end{eqnarray}
with torsion decoupling from the spinor, so that we can neglect it. The full energy with electrodynamics is
\begin{eqnarray}
&T^{\rho\sigma}\!\approx\!2\phi^{2}m\cos{\beta}u^{\sigma}u^{\rho}
\!+\!\frac{1}{4}F^{2}g^{\rho\sigma}\!-\!F^{\rho}_{\phantom{\rho}\alpha}F^{\sigma\alpha}
\end{eqnarray}
and because $\nabla_{\rho}T^{\rho\sigma}\!=\!0$ then 
\begin{eqnarray}
&q2\phi^{2} u_{\alpha}\!F^{\sigma\alpha}\!=\!m2\phi^{2} u^{\eta}\nabla_{\eta}(u^{\sigma}\cos{\beta})
\end{eqnarray}
having used Maxwell equations and the conservation of the electrodynamic current. Simplifying the module and employing again (\ref{momentum}) we eventually obtain
\begin{eqnarray}
&u^{\eta}\nabla_{\eta}P^{\sigma}\!=\!qF^{\sigma\alpha}u_{\alpha}
\label{Lorentz}
\end{eqnarray}
which is the Lorentz force in the Newton law.\! Notice that we have never used any assumption on the module being localized in order to get the macroscopic approximation, which means that in the present derivation all points and not only the peak of the matter distribution do follow the classical trajectory. In absence of electrodynamics
\begin{eqnarray}
&u^{\eta}\nabla_{\eta}P^{\sigma}\!=\!0
\label{Newton}
\end{eqnarray}
as the Newton law. Therefore the mass can be simplified, obtaining the equivalence principle, and if also $\beta\!\rightarrow\!0$ we have $u^{\eta}\nabla_{\eta}u^{\sigma}\!=\!0$ identically, which is merely the geodesic equation. Further, the entire derivation could have been obtained also without the macroscopic approximation, in which case we would obtain the particle trajectories with corrections due to the quantum potentials. The full form of the final expressions is too complicated to be insightful, but even without them one can already understand that they constitute the guiding equation.\! In fact, after having solved for $P^{\sigma}$ we can use
\begin{eqnarray}
P^{\sigma}\!=\!m\frac{d}{ds}x^{\sigma}
\end{eqnarray}
and solve for $x^{\sigma}\!=\!x^{\sigma}(s)$ giving the position of the particle in terms of the length parameter $s$ and which is therefore the trajectory of the particle, as in the dBB theory \cite{b}.

As we had already mentioned, the above derivation was based on no assumption regarding the localization of the matter distribution, which made it more general than the Ehrenfest theorem on the classical limit. However, it also allows a novel definition of particles that does not require them to be the manifestation of a localized module. And in fact in the dBB theory, particles are not the peak of the module but yet another entity that has to be postulated independently and which rides on the module according to the guiding equation. This interpretation, however, is a weak point of the dBB theory. In fact, in this case, the motion of a particle would be determined by the module, that is the wave function, which in principle depends on the configuration of all other particles. Whereas this link among all particles of the universe is non-local enough to ensure entanglement of all particles, this entanglement is mediated by the wave function, which is physical. Hence, any non-local behaviour is also physical. The possibility that acausal propagation may be observable creates some compatibility issue with relativity. While these problems might be circumvented by arguing that still no information is actually exchanged between distant systems, there seems to be no general consensus yet. Consequently, here we would like to take a different route, interpreting particles as the manifestation of a localized module, finding a different manner to explain correlation between states.

In the following, there are then two things we will have to do. One is to justify somehow how the module can be localized. The other is explaining how two states can be linked non-locally but in full compatibility with relativity.

The first of these two problems may be treated by considering that in full, the Dirac equations also contain the torsion of the space-time. Consider then the Dirac equations given in the following alternative form
\begin{eqnarray}
&\nabla_{\mu}\ln{\phi^{2}}\!-\!G_{\mu}\!+\!2ms_{\mu}\sin{\beta}\!=\!0\\
&\nabla_{\mu}\beta\!-\!2XW_{\mu}\!-\!K_{\mu}\!+\!2ms_{\mu}\cos{\beta}\!=\!0
\end{eqnarray}
with $G_{\mu}\!=\!-R_{\mu}\!+\!2P^{\rho}u^{\nu}s^{\alpha}\varepsilon_{\mu\rho\nu\alpha}$ and $K_{\mu}\!=\!-B_{\mu}\!+\!2P^{\iota}u_{[\iota}s_{\mu]}$ as yet another type of potentials. Via the straightforward manipulation of these equations, we can obtain
\begin{eqnarray}
\nonumber
&\nabla^{\mu}\left(\phi^{2}\nabla_{\mu}\beta\right)
\!-\!(8X^{2}M^{-2}\phi^{2}m\sin{\beta}+\\
&+2XW\!\cdot\!G\!+\!\nabla_{\mu}K^{\mu}\!+\!K_{\mu}G^{\mu})\phi^{2}\!=\!0\label{cont}\\
\nonumber
&\left|\nabla\beta/2\right|^{2}\!\!-\!m^{2}\!-\!\phi^{-1}\nabla^{2}\phi
\!+\!\frac{1}{2}(\nabla_{\mu}G^{\mu}+\\
&+\frac{1}{2}G^{2}\!-\!\frac{1}{2}K^{2}\!-\!2XW\!\cdot\!K\!-\!2X^{2}W^{2})\!=\!0\label{HJ}
\end{eqnarray}
the first being a continuity equation and the second being a Hamilton-Jacobi equation. Particularly interesting for us is the HJ equation for $\beta\!\rightarrow\!0$ because in this case
\begin{eqnarray}
\nonumber
&\nabla^{2}\phi\!+\!X^{2}W^{2}\phi\!+\!XW\!\cdot\!K\phi\!-\!\frac{1}{2}(\nabla_{\mu}G^{\mu}+\\
&+\frac{1}{2}G^{2}\!-\!\frac{1}{2}K^{2})\phi\!+\!m^{2}\phi\!=\!0
\end{eqnarray}
with $W_{\mu}$ left explicitly. As for the field equations for the propagating torsion field \cite{Hehl:1976kj}, they can be taken in their effective approximation $M^{2}W^{\mu}\!=\!2X\phi^{2}s^{\mu}$ which, upon a direct substitution, furnish the effective HJ equations
\begin{eqnarray}
\nonumber
&\nabla^{2}\phi\!-\!4X^{4}M^{-4}\phi^{5}\!+\!2X^{2}M^{-2}K\!\cdot\!s\phi^{3}-\\
&-\frac{1}{2}(\nabla_{\mu}G^{\mu}\!+\!\frac{1}{2}G^{2}\!-\!\frac{1}{2}K^{2}\!-\!2m^{2})\phi\!=\!0
\end{eqnarray}
which are now written in the form of Klein-Gordon equations for the module. They are non-linear with negative sign for the highest-order potential, which make them the candidate equations for a solitonic solution. Therefore, a localized module can be dynamically justified by torsion.

However, in practice, it is very difficult to actually find solutions of such non-linear field equations, though some approximated solutions can be found like those of \cite{Fabbri:2020ypd}, or those of \cite{Fabbri:2021weq} and \cite{Fabbri:2021nkn}. Either way, the solution can be seen as a localized but regular matter distribution in general.

To face the issue of entanglement, we begin by recalling some general features of the theory presented so far which may be of help. First of all, as it is well known, the Dirac equations contain the spinor field and its dynamical properties but also the tetradic fields. These tetrad fields are important for two reasons.\! A first is that without them we cannot write the spinor equation, highlighting how much the spinor fields are sensitive to the underlying structure of the background.\! Another is that tetrads contain more information as compared to the metric within the same background. In a given background of assigned metric, a basis of tetrads have a richer structure which can be felt by spinor fields. Secondly, both information about frame of reference and gravitational effects are generally found within tetrad fields, although only gravity can be found in the curvature tensor and henceforth determined by field equations with a source. So the information about pure geometry that can be found inside non-trivial tetrads in flat space-time remains undetermined. Genuine geometric effects in tetrads have no propagation, and no acausal behaviour can be imputed to them. Non-local actions are therefore not forbidden by any known physical principle.

In encoding what we can know about physics, tetrads complement the information contained in the spinor field and without being pre-determined.\! A flat space-time does not imply that pure geometric effects cannot be present, and in fact the tetrads can still be non-trivial, entering in the Dirac equations in a way that can have an impact on the spinor field. So in the following we will work out some consequences of a toy model based on an exact solution of the Dirac equations in a perfectly flat space-time.

Consider then the Minkowski metric, thus zero connection and flat space-time. We can write tetrads and spin connection as those found in \cite{Fabbri:2020ypd,Fabbri:2021weq,Fabbri:2021nkn}. Whatever its form, a solution is in general constituted by an assigned module and Yvon-Takabayashi angle that\! have to solve\! (\ref{dep1}-\ref{dep2}) in a specific background that is given. Equivalently, we can also write the spinor field according to the form
\begin{eqnarray}
&\!\psi\!=\!\phi e^{-\frac{i}{2}\beta\boldsymbol{\pi}}
e^{-iq\alpha}\left(\begin{tabular}{c}
$1$\\
$0$\\
$1$\\
$0$
\end{tabular}\right)
\label{1}
\end{eqnarray}
solving (\ref{D}) for a specific set of tetradic fields that is also given as background. These are general results \cite{Fabbri:2020ypd,Fabbri:2021weq}. For this solution, however, it is also possible to assign a very special alternative form that is given for the same module and Yvon-Takabayashi angle. Quite simply it is 
\begin{eqnarray}
&\!\psi\!=\!\phi e^{-\frac{i}{2}\beta\boldsymbol{\pi}}
e^{-iq\alpha}\left(\begin{tabular}{c}
$0$\\
$1$\\
$0$\\
$1$
\end{tabular}\right)
\label{2}
\end{eqnarray}
corresponding to the very same material distribution but with an opposite spin. This is not a surprise because it is well known that spinors have two basic spin orientations, as wanted by the Pauli principle. The two solutions above differ from (\ref{spinorch}) for the fact that they have been taken in their rest frame and spin aligned along the third axis, as also customary. But nonetheless, one might wonder what additional information could be encoded within the $\boldsymbol{L}^{-1}$ matrix. To keep things simple, we will still remain in the rest frame. But there we consider a rotation of the form
\begin{eqnarray}
&\!\!\!\!\!\!\!\!\boldsymbol{L}^{-1}\!=\!\left(\begin{array}{cccc}
\!\cos{\zeta/2}\!&\!\sin{\zeta/2}\!&\!0\!&\!0\\
\!-\sin{\zeta/2}\!&\!\cos{\zeta/2}\!&\!0\!&\!0\\
\!0\!&\!0\!&\!\cos{\zeta/2}\!&\!\sin{\zeta/2}\\ 
\!0\!&\!0\!&\!-\sin{\zeta/2}\!&\!\cos{\zeta/2}
\end{array}\right)\label{ROTATION}
\end{eqnarray}
with $\zeta\!=\!\omega t$ and $\omega$ constant. The appearance of such new term determines the appearance of an additional
\begin{eqnarray}
\boldsymbol{L}^{-1}\partial_{t}\boldsymbol{L}\!=\!-\omega\boldsymbol{\sigma}_{13}
\end{eqnarray}
so that (\ref{LdL}) yields
\begin{eqnarray}
\partial_{t}\xi_{13}\!=\!-\omega
\end{eqnarray}
as Goldstone mode of this state. Because of (\ref{R}) we have that $\Omega_{13t}\!=\!\partial_{t}\xi_{13}$ and consequently we get
\begin{eqnarray}
&\Omega_{13t}\!=\!-\omega
\end{eqnarray}
as additional component of the spin connection. As it is clear, there is no contribution to the curvature, for which we still have flatness. The general form of the spinor (\ref{spinorch}) in both the above cases (\ref{1}-\ref{2}) is therefore
\begin{eqnarray}
&\!\psi\!=\!\phi e^{-\frac{i}{2}\beta\boldsymbol{\pi}}
e^{-iq\alpha}\left(\begin{tabular}{c}
$\cos{\zeta/2}$\\
$-\sin{\zeta/2}$\\
$\cos{\zeta/2}$\\
$-\sin{\zeta/2}$
\end{tabular}\right)
\label{s1}
\end{eqnarray}
having $s^{3}\!=\!\cos{\zeta}$ alongside to
\begin{eqnarray}
&\!\psi\!=\!\phi e^{-\frac{i}{2}\beta\boldsymbol{\pi}}
e^{-iq\alpha}\left(\begin{tabular}{c}
$\sin{\zeta/2}$\\
$\cos{\zeta/2}$\\
$\sin{\zeta/2}$\\
$\cos{\zeta/2}$
\end{tabular}\right)
\label{s2}
\end{eqnarray}
with $s^{3}\!=\!-\cos{\zeta}$ therefore showing the opposition of the two spin orientations. Then, while maintaining opposite orientation, both spins display a flipping that depends on $\zeta$ over time. The Goldstone state has no dependence on spatial coordinates and it will remain the same even when the two solutions have space-like distance. Now, suppose that a measurement be performed on the first solution so to force it to collapse onto the state\! of definite spin.\! If we perform a measurement fixing the first spinor to its form
\begin{eqnarray}
&\!\psi\!=\!\phi e^{-\frac{i}{2}\beta\boldsymbol{\pi}}
e^{-iq\alpha}\left(\begin{tabular}{c}
$1$\\
$0$\\
$1$\\
$0$
\end{tabular}\right)
\end{eqnarray}
then $s^{3}\!=\!1$ hence showing that the spin is in the up configuration. Since this state would still have to be solution of the Dirac equation, the spin connection collapses onto the case in which we have that $\omega\!=\!0$ and because this is a constant then it will remain in this state, so that
\begin{eqnarray}
&\Omega_{13t}\!=\!0
\end{eqnarray}
and since the spin connection is uniquely defined as background then this must also be the value of the spin connection of the second state. As we want this state to still be a solution of the Dirac equation, then
\begin{eqnarray}
&\!\psi\!=\!\phi e^{-\frac{i}{2}\beta\boldsymbol{\pi}}
e^{-iq\alpha}\left(\begin{tabular}{c}
$0$\\
$1$\\
$0$\\
$1$
\end{tabular}\right)
\end{eqnarray}
with $s^{3}\!=\!-1$ and so that the spin is now in the down configuration. Summarizing, forcing the first solution into a spin-up state implies, through the $\omega\!=\!0$ condition,\! that a spin-down state be fixed for the second solution, and the full process can take place no matter how distant are the two solutions. Notice that such process would have been exactly the same if we had the first solution collapse onto the spin-down state and the second solution collapse onto the spin-up state. This uniform spin flip guarantees lack of pre-determination in spin orientation, and thus results are statistically distributed as it is necessary in quantum mechanics. Yet, a measurement fixing one spin also fixes the other spin, and it does so immediately. This process is mediated by the spin connection, and in particular by the component that arises as Goldstone state $\boldsymbol{L}^{-1}\partial_{\nu}\boldsymbol{L}$ in the structure of the spinor field. This degree of freedom does not\! encode physical\! interactions since\! it gives rise to no contribution in the curvature, and therefore it can not be determined by any field equation. So, the information that is transferred between the two spinors through their common Goldstone state is not restricted to be causal as it does not have any propagation in the first place.\! Hence compatibility with the principles of relativity is obvious.

As an example, let us next try to apply such a concept for a specific solution, that is that of \cite{Fabbri:2020ypd}, of which we will consider only the exterior branch. We will have two wave functions with an opposite spin orientation. And we will work in spherical coordinates for compactness. Of these two opposite-spin wave functions, the first that we shall consider is given according to the following expression
\begin{eqnarray}
&\!\!\!\!\psi\!=\!\frac{K}{r\sqrt{\sin{\theta}}}e^{-r\sqrt{\varepsilon(2m-\varepsilon)}}
e^{-it(m-\varepsilon)}\left(\begin{tabular}{c}
$1$\\
$0$\\
$1$\\
$0$
\end{tabular}\right)\label{exu1}
\end{eqnarray}
with tetrads
\begin{eqnarray}
&\!\!\!\!e_{0}^{t}\!=\!\cosh{\alpha}\ \ \ \ e_{2}^{t}\!=\!-\sinh{\alpha}\label{exu2}\\
&\!\!\!\!e_{1}^{r}\!=\!-1\label{exu3}\\
&e_{3}^{\theta}\!=\!\frac{1}{r}\label{exu4}\\
&\!\!\!\!e_{0}^{\varphi}\!=\!-\frac{1}{r\sin{\theta}}\sinh{\alpha}\ \ \ \ 
e_{2}^{\varphi}\!=\!\frac{1}{r\sin{\theta}}\cosh{\alpha}\label{exu5}
\end{eqnarray}
giving spin connection
\begin{eqnarray}
&\Omega_{13\theta}\!=\!-1\label{exu6}\\
&\Omega_{01\varphi}\!=\!-\sin{\theta}\sinh{\alpha}\label{exu7}\\ 
&\Omega_{03\varphi}\!=\!\cos{\theta}\sinh{\alpha}\label{exu8}\\
&\Omega_{12\varphi}\!=\!-\sin{\theta}\cosh{\alpha}\label{exu9}\\
&\Omega_{23\varphi}\!=\!-\cos{\theta}\cosh{\alpha}\label{exu10}
\end{eqnarray}
where $\sinh{\alpha}\!=\!\sqrt{\varepsilon(2m-\varepsilon)}/(m\!-\!\varepsilon)$ with $m\!>\!\varepsilon\!>\!0$ and $K$ a generic constant. This corresponds to the spin-up case and it is a solution of the Dirac equations. Similarly it is possible to consider the alternative wave function
\begin{eqnarray}
&\!\!\!\!\psi\!=\!\frac{K}{r\sqrt{\sin{\theta}}}e^{-r\sqrt{\varepsilon(2m-\varepsilon)}}
e^{-it(m-\varepsilon)}\left(\begin{tabular}{c}
$0$\\
$1$\\
$0$\\
$1$
\end{tabular}\right)\label{exd1}
\end{eqnarray}
with tetrads
\begin{eqnarray}
&\!\!\!\!e_{0}^{t}\!=\!\cosh{\alpha}\ \ \ \ e_{2}^{t}\!=\!-\sinh{\alpha}\label{exd2}\\
&\!\!\!\!e_{1}^{r}\!=\!1\label{exd3}\\
&e_{3}^{\theta}\!=\!-\frac{1}{r}\label{exd4}\\
&\!\!\!\!e_{0}^{\varphi}\!=\!-\frac{1}{r\sin{\theta}}\sinh{\alpha}\ \ \ \ 
e_{2}^{\varphi}\!=\!\frac{1}{r\sin{\theta}}\cosh{\alpha}\label{exd5}
\end{eqnarray}
giving spin connection
\begin{eqnarray}
&\Omega_{13\theta}\!=\!-1\label{exd6}\\
&\Omega_{01\varphi}\!=\!\sin{\theta}\sinh{\alpha}\label{exd7}\\ 
&\Omega_{03\varphi}\!=\!-\cos{\theta}\sinh{\alpha}\label{exd8}\\
&\Omega_{12\varphi}\!=\!\sin{\theta}\cosh{\alpha}\label{exd9}\\
&\Omega_{23\varphi}\!=\!\cos{\theta}\cosh{\alpha}\label{exd10}
\end{eqnarray}
where $\sinh{\alpha}\!=\!\sqrt{\varepsilon(2m-\varepsilon)}/(m\!-\!\varepsilon)$ with $m\!>\!\varepsilon\!>\!0$ and $K$ generic constant. This corresponds to the spin-down case and it is a solution of the Dirac equations. As easy to see these solutions are square-integrable (albeit their energy has a logarithmic divergence near the origin of the radial coordinate). By applying now the rotation (\ref{ROTATION}) we will get that the first spinor becomes of the form (\ref{s1}) as
\begin{eqnarray}
&\!\!\!\!\psi\!=\!\frac{K}{r\sqrt{\sin{\theta}}}e^{-r\sqrt{\varepsilon(2m-\varepsilon)}}
e^{-it(m-\varepsilon)}\left(\begin{tabular}{c}
$\cos{\zeta/2}$\\
$-\sin{\zeta/2}$\\
$\cos{\zeta/2}$\\
$-\sin{\zeta/2}$
\end{tabular}\right)\label{exu1t}
\end{eqnarray}
with the real representation of (\ref{ROTATION}) inducing the corresponding rotation on the tetrads
\begin{eqnarray}
&\!\!\!\!e_{0}^{t}\!=\!\cosh{\alpha}\ \ \ \ 
e_{2}^{t}\!=\!-\sinh{\alpha}\label{exu2t}\\
&\!\!\!\!e_{1}^{r}\!=\!-\cos{\zeta}\ \ \ \ 
e_{3}^{r}\!=\!-\sin{\zeta}\label{exu3t}\\
&\!\!\!\!e_{1}^{\theta}\!=\!-\frac{1}{r}\sin{\zeta}\ \ \ \ 
e_{3}^{\theta}\!=\!\frac{1}{r}\cos{\zeta}\label{exu4t}\\
&\!\!\!\!e_{0}^{\varphi}\!=\!-\frac{1}{r\sin{\theta}}\sinh{\alpha}\ \ \ \ 
e_{2}^{\varphi}\!=\!\frac{1}{r\sin{\theta}}\cosh{\alpha}\label{exu5t}
\end{eqnarray}
and hence on the spin connection
\begin{eqnarray}
&\Omega_{13t}\!=\!-\omega\label{exu6t}\\
&\Omega_{13\theta}\!=\!-1\label{exu7t}\\
&\Omega_{01\varphi}\!=\!-\sin{(\theta\!+\!\zeta)}\sinh{\alpha}\label{exu8t}\\ 
&\Omega_{03\varphi}\!=\!\cos{(\theta\!+\!\zeta)}\sinh{\alpha}\label{exu9t}\\
&\Omega_{12\varphi}\!=\!-\sin{(\theta\!+\!\zeta)}\cosh{\alpha}\label{exu10t}\\
&\Omega_{23\varphi}\!=\!-\cos{(\theta\!+\!\zeta)}\cosh{\alpha}\label{exu11t}
\end{eqnarray}
while the second spinor becomes of the form (\ref{s2}) as
\begin{eqnarray}
&\!\!\!\!\psi\!=\!\frac{K}{r\sqrt{\sin{\theta}}}e^{-r\sqrt{\varepsilon(2m-\varepsilon)}}
e^{-it(m-\varepsilon)}\left(\begin{tabular}{c}
$\sin{\zeta/2}$\\
$\cos{\zeta/2}$\\
$\sin{\zeta/2}$\\
$\cos{\zeta/2}$
\end{tabular}\right)\label{exd1t}
\end{eqnarray}
with the real representation inducing the corresponding rotation on the tetrads
\begin{eqnarray}
&\!\!\!\!e_{0}^{t}\!=\!\cosh{\alpha}\ \ \ \ 
e_{2}^{t}\!=\!-\sinh{\alpha}\label{exd2t}\\
&\!\!\!\!e_{1}^{r}\!=\!\cos{\zeta}\ \ \ \ 
e_{3}^{r}\!=\!\sin{\zeta}\label{exd3t}\\
&\!\!\!\!e_{1}^{\theta}\!=\!\frac{1}{r}\sin{\zeta}\ \ \ \ 
e_{3}^{\theta}\!=\!-\frac{1}{r}\cos{\zeta}\label{ex41t}\\
&\!\!\!\!e_{0}^{\varphi}\!=\!-\frac{1}{r\sin{\theta}}\sinh{\alpha}\ \ \ \ 
e_{2}^{\varphi}\!=\!\frac{1}{r\sin{\theta}}\cosh{\alpha}\label{exd5t}
\end{eqnarray}
and hence on the spin connection
\begin{eqnarray}
&\Omega_{13t}\!=\!-\omega\label{exd6t}\\
&\Omega_{13\theta}\!=\!-1\label{exd7t}\\
&\Omega_{01\varphi}\!=\!\sin{(\theta\!+\!\zeta)}\sinh{\alpha}\label{exd8t}\\ 
&\Omega_{03\varphi}\!=\!-\cos{(\theta\!+\!\zeta)}\sinh{\alpha}\label{exd9t}\\
&\Omega_{12\varphi}\!=\!\sin{(\theta\!+\!\zeta)}\cosh{\alpha}\label{exd10t}\\
&\Omega_{23\varphi}\!=\!\cos{(\theta\!+\!\zeta)}\cosh{\alpha}\label{exd11t}
\end{eqnarray}
and they are both solutions of the Dirac equations. Hence we see that\! both wave functions display the above uniform rotation, with the spin connection that has generated the additional component $\Omega_{13t}\!=\!-\omega$ exactly as we discussed above.\! Notice that\! such a component does not depend on the variables of the system but remark also that it is not an absolute constant. The independence on the position of the particle means that the dynamics will remain the same even if the two particles were separated. However, any observation breaking the rotation by fixing $\omega\!=\!0$ will have the effect of producing the collapse of both spinors simultaneously. In fact, suppose that an observation were performed at a time for which more or less $\omega t\!=\!2n\pi$ then solution (\ref{exu1t}-\ref{exu11t}) would be (\ref{exu1}-\ref{exu10}) plus the $\Omega_{13t}\!=\!-\omega$ condition and (\ref{exd1t}-\ref{exd11t}) as (\ref{exd1}-\ref{exd10}) plus the $\Omega_{13t}\!=\!-\omega$ condition. If now the system were disturbed so that $\omega\!=\!0$ then the rotation would stop, simultaneously locking the first solution to the spin-up state and the second solution to the spin-down state.\! If we had about $\omega t\!=\!2n\pi\!+\!\pi$ then solution (\ref{exu1t}-\ref{exu11t}) would be (\ref{exd1}-\ref{exd10}) plus the $\Omega_{13t}\!=\!-\omega$ condition and (\ref{exd1t}-\ref{exd11t}) as (\ref{exu1}-\ref{exu10}) plus the $\Omega_{13t}\!=\!-\omega$ condition. If now the system were disturbed so that $\omega\!=\!0$ then the rotation would stop, simultaneously locking the first solution to spin-down states and the second solution to spin-up states. This is what we had discussed above.

Contrary to the dBB interpretation, where, as already mentioned, entanglement is due to observable degrees of freedom, and thus non-local effects are real, here the correlation of two observables occurs through the Goldstone degrees of freedom, which have no local restriction given that their propagation is not restricted by anything.\! With the original terminology \cite{epr,ab}, \cite{B} we may say that in the dBB interpretation non-local hidden variables are the positions of the particles, while here the non-local hidden variables are the Goldstone state of the spinorial field.

The single measurement is also completely determined through the knowledge of the parameter $\zeta\!=\!\omega t$ and ultimately on $t$ but this requires the knowledge of the initial time $t_{0}$ as boundary condition. Knowledge of this boundary condition is therefore the condition in terms of which of all possible states only a special state is selected hence entailing a form of symmetry breaking. It is of particular type as it occurs for a special state. And again it occurs only for that special state due to boundary conditions.

Analogies with the previous case are found in the fact that both types of symmetry breaking are specific to one given wave function. Differently from the previous case, where\! symmetry breaking meant\! choosing one solution of many, here symmetry breaking means choosing a specific observable property for a given assigned solution.
%%%%%%%%%%%%%%%%%%%%%%%%%%%%%%%%%%%%%%%%%%%%%%%%%%%%%%%%%%%%%%%%%%%%%%%%%%%%%%%%%%%%%%%%%%%%%%%%%%%
%%%%%%%%%%%%%%%%%%%%%%%%%%%%%%%%%%%%%%%%%%%%%%%%%%%%%%%%%%%%%%%%%%%%%%%%%%%%%%%%%%%%%%%%%%%%%%%%%%%
\section{Conclusion}
In this paper, we have considered symmetry breaking occurring in two situations, universally and particularly, and we have discussed these ideas in terms of three possible examples. The first was about cosmological effects of the standard model of particle physics.\! We recalled the way in which symmetry gets broken by the Higgs vacuum and by the choice of the specific gauge field configuration selected by the Higgs isospin. Since these selections occur throughout the universe, symmetry breaking is universal.

The second case was a situation that was formally analogous to the one of the Higgs field. We have shown how, similarly to the Higgs field, general spinor fields are written in polar form. For them we identified the Goldstone states and shown that they are absorbed as longitudinal components in the $P_{\alpha}$ and $R_{ij\alpha}$ tensors. So we have seen that there arises a specific spin axial-vector and we have recalled how its direction selects a specific spinor field as solution to the Dirac equations. Because the selection of this solution is due to boundary conditions that are valid for such a case solely, symmetry breaking is particular.

The third case was an entirely different situation, that was regarding the process of measurement within general quantum mechanical systems. After having presented the version of the dBB formalism that was written in a wholly relativistic form and in presence of spin, we had shown a toy model in which a pair of spin-up and spin-down states were found to possess a uniform rotation that could have been maintained over large distances but which could also be made to collapse for one spin state instantly forcing a collapse of the other spin state. This toy model for pairs of entangled spins does not have compatibility problems with relativity because in it the correlation of two states is ensured by the component of the spin connection arising from the Goldstone degrees of freedom of the spinor field, which have a peculiar property. They are given by the term $\boldsymbol{L}^{-1}\partial_{\nu}\boldsymbol{L}$ and as such they do not contribute to the curvature tensor, showing that they cannot carry any gravitational information but only frame-related types of information in general. As the gravitational information would go into the curvature and as such it would have to verify Einstein equations ensuring the causal propagation of all gravitational\! degrees of freedom,\! information about frames does not go in the curvature and so for it there is no field equation restricting its propagation.\! Whereas an interaction mediated by some physical field would have to respect physical locality, entanglement as described by frames does not have to obey constraints. In other words, even if locality must be ensured for all fields that are the solutions of field equations, not all fields are solutions of field equations. There might be non-local objects even in a full relativistic environment.\! And thus employing them to have a description of a non-local action is compatible with relativity. In our toy model, they are the Goldstone states of\! spinor fields,\! and they\! are recognized as non-local hidden variables. These are fixed by boundary conditions such as the time $t_{0}$ that make the wave function collapse onto a single state. As the boundary conditions pick only one state, the resulting symmetry breaking is particular.

To conclude this discussion about symmetry breaking types, we would like to stress that the example of universal type and the first example of particular type seem to be more alike than the two examples of particular type, for the following reasons. The example of universal type and the first example of particular type are different for the fact that in the former the symmetry group is among different fields while in the latter the symmetry group is among different components of the same field, but apart from this they are analogous in any aspect, even formally.

The two examples of particular type are different in the fact that while in the former the choice of the boundary conditions is made for $\beta$ and $\phi^{2}$ in the latter the choice of the boundary conditions is made for the Goldstone states of the spinor fields. While the Dirac equations specify the physical properties of the wave function, no equation can determine the propagation of its Goldstone states. Hence one type of boundary conditions is the usual type needed to specify the solution of a given field equation whereas the other type of boundary conditions is a new type that specifies an observable property of a given solution.

In our toy model, the property in exam was the orientation of the spin axial-vector, which cannot be determined by field equations and yet it is necessary to establish the results of spin observations. This analysis points toward the importance of Goldstone states of a spinor field in encoding non-directly observable information.
%%%%%%%%%%%%%%%%%%%%%%%%%%%%%%%%%%%%%%%%%%%%%%%%%%%%%%%%%%%%%%%%%%%%%%%%%%%%%%%%%%%%%%%%%%%%%%%%%%%
%%%%%%%%%%%%%%%%%%%%%%%%%%%%%%%%%%%%%%%%%%%%%%%%%%%%%%%%%%%%%%%%%%%%%%%%%%%%%%%%%%%%%%%%%%%%%%%%%%%

%%%%%%%%%%%%%%%%%%%%%%%%%%%%%%%%%%%%%%%%%%%%%%%%%%%%%%%%%%%%%%%%%%%%%%%%%%%%%%%%%%%%%%%%%%%%%%%%%%%
\end{document}